\documentclass[aps, prl, 10pt, twocolumn, superscriptaddress]{revtex4-2}
\usepackage{silence}
\usepackage{url}
\usepackage{comment}
\usepackage[normalem]{ulem}
\usepackage{graphicx}
\usepackage[colorlinks=true, linkcolor=blue, citecolor=blue, urlcolor=blue]{hyperref}
\usepackage{orcidlink}

\begin{document}

\title{On-Demand Millisecond Storage of Spectrotemporal Modes at Telecom Wavelength}

\author{Anuj Sethia\,\orcidlink{0000-0002-7140-3662}}
\author{Nasser Gohari Kamel\,\orcidlink{0009-0003-7869-2561}}
\author{Daniel Oblak\,\orcidlink{0000-0002-0277-3360}}

\email[Corresponding email: ]{doblak@ucalgary.ca}
\affiliation{Department of Physics and Astronomy, University of Calgary, Calgary, Alberta T2N 1N4, Canada.}
\affiliation{Institute of Quantum Science and Technology, University of Calgary, Calgary, Alberta T2N 1N4, Canada.}

\date{\today}

\begin{abstract}
The realization of scalable quantum networks for distribution of entanglement over long distances hinges on quantum repeaters.
To outperform the exponential transmission loss in optical fibers, quantum repeaters must employ multiplexing schemes in the temporal, spectral, or spatial domain. 
The performance of such a multiplexed scheme is contingent on efficient quantum memories offering both extended storage times and large multimode capacities.
Toward this goal, we experimentally demonstrate a memory platform operating at telecom wavelength using an Er\textsuperscript{3+}:Y\textsubscript{2}SiO\textsubscript{5} crystal. 
We record on-demand storage and recall of weak coherent pulses for up to $1$\,ms, exceeding by an order of magnitude that of previously reported demonstrations based on Er\textsuperscript{3+}. 
The memory exhibits an efficiency of 10.36\% at 300\,\textmu s storage time with a signal-to-noise ratio of $10.9$, mainly limited by spontaneous emission noise. 
We also store two spectral modes representing a frequency-bin encoding and retrieve the state with an overlap of 81.8\%.
Furthermore, we demonstrate multimode capacity by simultaneously storing 20 temporal and 3 spectral modes  with on-demand and selective recall capabilities, serving as a promising step toward a scalable quantum repeater architecture.

\end{abstract}

\maketitle

Future quantum networks~\cite{QInternet_kimble-2008, QInterenet_wehner-2018} are envisioned to enable transformative applications, such as distributed quantum computing~\cite{QComp_Distributed_cuomo-2020, QComp_Blind_giovannetti-2013}, quantum-enhanced distributed sensing~\cite{QSensing_ge-2018, QSensing_proctor-2018, QSensing_zhang-2020}, precise clock distribution~\cite{Clocks_QNetwork_komar-2014}, and secure communication~\cite{QKD_QNet_zhang-2018}. 
Since exponential photon loss in optical fibers limits the communication link distances, quantum repeaters (QRs)~\cite{QRepeater_simon-2007, QRepeater_sangouard-2011} are essential to implement such long-range quantum networks.
To achieve practical entanglement distribution rates, most QR protocols require high-performance quantum memories (QMs)~\cite{QMem_Review_tittel-2009, QMem_Review_simon-2010, QMem_Review_wei-2022}. 
Specifically, QMs must offer a wide bandwidth, high multimode capacity, high storage efficiency, and storage times on the order of milliseconds, equivalent to the round trip time over the link~\cite{QMem_Review_wei-2022, QMem_Review_zhou-2023}.
The development of such high-performance QMs requires both the exploration of new material platforms and the advancement of implementation protocols.

Rare-earth ion-doped crystals are a leading candidate for QMs due to their long coherence lifetimes~\cite{REI_Coherence_ma-2021, REI_Coherence_ortu-2022, REI_Coherence_wang-2025}, along with their proven multimode capacity~\cite{REI_AFC_afzelius-2009, REI_Multimode_sinclair-2014,REI_Multimode_jobez-2016, ErFiber_wei-2024, REI_Multimode_lago-rivera-2021}, wide bandwidths~\cite{QMem_Review_wei-2022, QMem_Review_zhou-2023}, and potential for near-unity storage efficiency~\cite{REI_Cavity_afzelius-2010, REI_Cavity_duranti-2024}. 
Among these, erbium ions (Er\textsuperscript{3+}) are particularly compelling because their optical transition falls within the telecom C band. 
This critical property enables Er\textsuperscript{3+}-based QM to interface directly with existing low-loss telecommunication fiber infrastructure, positioning them as an ideal platform for scalable, long-distance quantum networks. 
Such QMs have been successfully demonstrated across a range of host platforms, including optical fibers~\cite{ErFiber_jin-2015, ErFiber_saglamyurek-2015, ErFiber_puigibert-2020, ErFiber_wei-2024, ErFiber_Nasser_2025}, LiNbO\textsubscript{3}~\cite{ErLN_AFC_Ent_askarani-2019, ErLN_Waveguide_zhang-2023}, and Y\textsubscript{2}SiO\textsubscript{5}~\cite{ErYSO_ROSE_minnegaliev-2023, ErYSO_EntStorage_jiang-2023, ErYSO_AFC_Yasui2024, ErYSO_Multidimentional_Li2024, ErYSO_Teleportation_An2025}, with additional promising platforms continuously being explored~\cite{ErCaWo4_dantec-2021, ErTiO2_gupta-2025}. 
Although host platforms such as Y\textsubscript{2}SiO\textsubscript{5} exhibit coherence times up to $4$\,ms~\cite{ErYSO_Magnetic_gtensor_Sun2008, ErYSO_MagneticField_bottger-2009}, the longest storage time demonstrated to date is only 110\,\textmu s~\cite{ErYSO_Rose_Dajczgewand2015}. 
Achieving practical millisecond-scale storage along with multimode capacity, as required for QRs, therefore, remains an open challenge.

An important challenge limiting longer storage times arises from the QM protocol itself. The atomic frequency comb (AFC)-based~\cite{REI_AFC_afzelius-2009} schemes face inherent constraints due to the difficulty of preparing sufficiently narrow comb structures~\cite{QMem_NarrowAFC_businger-2022, QMem_NarrowAFC_sanchez-2025}. 
Furthermore, spin-wave storage, which would enable both extended storage times and on-demand recall, has not yet been demonstrated in Er\textsuperscript{3+} based systems, due to the complexity of resolving the spin transitions within the inhomogeneous linewidth~\cite{ErYSO_Spectro_Stuart2021, ErYSO_B7T_Cohrenece_Rancic2018, ErYSO_Spectro_Kukharchyk2021, ErYSO_Zefoz_matsuura-2026}. 
The revival of silenced echo (ROSE) protocol~\cite{Rose_Intro_damon-2011, Rose_RAP_de-seze-2005, Rose_RAP_pascual-winter-2013} offers long storage time and on-demand recall capabilities; however, its reliance on spatial mismatch of optical fields makes it difficult to implement on a scalable integrated platform.

In this letter, we first identify the performance requirements a QM must meet to enhance entanglement distribution in a QR compared to direct fiber transmission.
To address these requirements, we employ the chirped pulse phase encoding (CPPE) protocol~\cite{QMem_CPPI_morton-2022, QMem_CPPI_kamel-2025}, which is closely related to ROSE~\cite{Rose_Intro_damon-2011}, to evaluate a telecom-wavelength storage platform using Er\textsuperscript{3+} ions doped in Y\textsubscript{2}SiO\textsubscript{5}. 
We achieve on-demand storage and recall of weak coherent pulses with an input mean photon number $\approx 720$ for durations ranging from 300\,\textmu s up to 1\,ms, with efficiencies of 10.36\% and 1.42\%, respectively.
We also realize a multimode capacity comprising 20 temporal and 3 spectral modes at a storage time of 800\,\textmu s, representing a significant step toward practical QR implementations.
We finally evaluate the performance of the system by estimating the overlap between the stored and retrieved states in the $Z$ and $X$ bases within the frequency-bin encoding framework.

\begin{figure*}[t!]
    \centering
    \includegraphics[width = \linewidth]{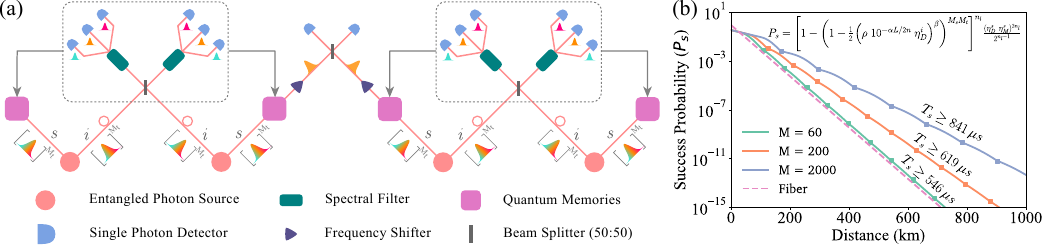}
    \caption{(a) Illustration of a spectrotemporal multiplexed quantum repeater with two elementary links. The protocol begins with the generation of $M_t$ entangled photon pairs simultaneously across $M_s = 3$ spectral modes using probabilistic nonlinear sources. The signal photons (\textit{s}) are stored in an on-demand QM, while the idler photons (\textit{i}) travel through the channel for entanglement swapping. Idler photons from adjacent nodes interfere on a 50:50 beam splitter and are spectrally resolved for individual detection, as showcased in Ref. \cite{ QRepeater_Multimode_chakraborty-2025}. A successful BSM triggers a feedback signal to recall the corresponding spectral mode from the QM, followed by an appropriate frequency shift for the subsequent swapping between signal photons of neighboring links.
    (b) Entanglement distribution success probability $P_s$ as a function of distance. The solid curves show the performance of a quantum repeater for different total multimode capacity $M = M_t \times M_s$ given memory parameters $\eta_o=65\%$ and $T_2=3$\,ms. The pink dashed curve represents direct transmission through SMF-28 fiber. The success probability is optimized for the number of elementary links ($n_l$), and the squares represent an increment in $n_l$ starting from $n_l=1$. Additionally, the minimum QM storage time ($T_S$) required for each scenario is estimated. See Supplemental Material (SM)~\cite{SM} for details on simulation parameters.}
    \label{fig:Qrepeater}
\end{figure*}

\textit{Multiplexed quantum repeater—}The overall entanglement distribution rate over the elementary links of a QR is constrained by the probabilistic nature of entangled-pair generation and fiber transmission, nonunit detector efficiencies, the inherent limitations of Bell-state measurements (BSM) with linear optics, and finite QM efficiency.
Fortunately, these limitations can be overcome by multiplexing a large number of modes in various degrees of freedom, such as time, frequency, or space~\cite{REI_Multimode_sinclair-2014, REI_Multimode_jobez-2016, REI_Multimode_lago-rivera-2021, QRepeater_Multimode_chakraborty-2025}. 
By attempting entanglement distribution across many modes in parallel, the overall success probability of entanglement swapping over all the elementary links can approach unity. 
However, implementing such a scheme requires a multimode QM capable of storing as well as selectively retrieving a large number of modes.

If provided with a QM featuring both spectral and temporal multimode capacity, one may incorporate it in the multiplexed QR scheme illustrated in Fig.~\ref{fig:Qrepeater}(a). The focus on spectral multiplexing, in addition to temporal multiplexing, is due to its compatibility with common broadband entangled photon pair sources, e.g., those based on spontaneous parametric down-conversion.
A key challenge in typical spectral multiplexing schemes is, however, the need for resolving the different spectral modes before detection at each node~\cite{REI_Multimode_sinclair-2014}. 
Our QR architecture overcomes this obstacle significantly by leveraging the QM's inherent selective recall capabilities, thus substantially reducing the architecture complexity.

QMs with high multimode capacities have been demonstrated~\cite{ErFiber_wei-2024, QMem_NarrowAFC_businger-2022}, but have been limited by short, fixed storage times and low retrieval efficiencies, which restrict their usefulness in long-distance quantum networks. To systematically identify the memory specifications required for a QR to surpass direct transmission using the same photon-pair source, we estimate the entanglement distribution success probability $P_s$ as a function of distance as illustrated in Fig.\ref{fig:Qrepeater}(b) (see SM \cite{SM} for more details). The storage efficiency of the memory is suitably fitted by an exponential decay as $\eta_{M}^{\textit{s}}(T_S) = \eta_o \exp \left[-4T_S/T_2\right]$, where $\eta_o$ is the memory efficiency in the limit of zero storage time and $T_2$ the memory coherence time. On the other hand, the round trip travel of signals between the memory and the BSM over an elementary link prescribes the required storage time in a QR as $T_S=L/(n_l\,v)$, where $L$ is the total link distance, $n_l$ is the optimal number of elementary links, and $v$ is the speed of light in glass. Based on these dependencies, the success probability $P_s$ is plotted for different memory parameters in Fig.~\ref{fig:Qrepeater}(b), along with the success probability for direct transmission. Of course, direct transmission may feature fewer limitations on pulse bandwidth and, thus, realistically allow higher repetition rates $\nu$. In this case, more stringent requirements for memory performance must be met, for the overall entanglement distribution rate $\nu\times P_s$ to exceed that for direct transmission.

\begin{figure*}[ht!]
    \includegraphics[width = \linewidth]{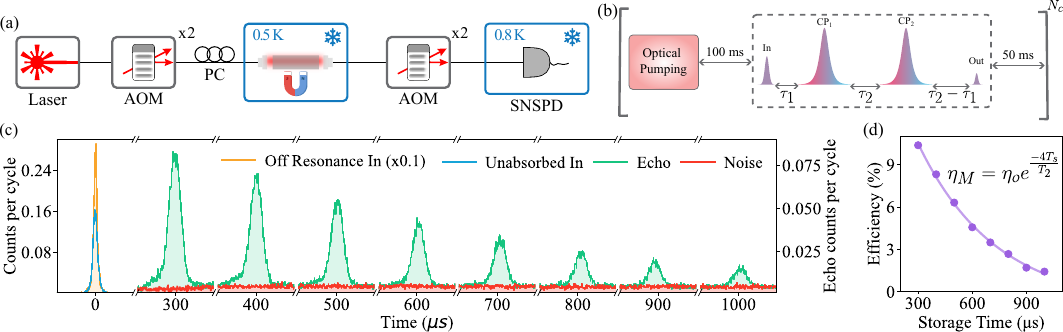}
    \caption{(a) Schematic of the experimental setup. The two acousto-optic modulators (AOMs) before the crystal prepare input modes and the CPs. The AOMs after the crystal, with combined efficiency of $\eta_\mathrm{AOM}=35\%$, allow only the input and desired echo to pass through for detection via superconducting nanowire single photon detectors SNSPD, featuring $\eta_\mathrm{D}=67\%$ detection efficiency. 
    (b) Illustration of the pulse sequence for a storage cycle of the CPPE memory protocol. The storage and recall of input modes is performed using CP\textsubscript{1} and CP\textsubscript{2}, as shown in the dashed box. 
    (c) Timing histogram of photon counts from SNSPD for $N_c = 9000$ cycles. The leftmost panel showcases the off-resonance input, with an estimated mean photon number of 720 before the crystal, and the unabsorbed input during the memory sequence. 
    Other panels display the recalled echo and the noise (measured without the input) for experiments with different storage times. The SNR of the recalled echo is 10.90 for 300\,\textmu s and 1.52 for $1$\,ms.
    (d) The experimental memory retrieval efficiency, estimated after subtracting the noise floor, is plotted against storage time and fitted with an exponential decay~\cite{REI_Multimode_lago-rivera-2021}. The fit's decay constant yields $\eta_o = 23.05\%$ and coherence time, $T_2 = 858.4 \pm 80.4$\,\textmu s, consistent with the two-pulse photon-echo measurement~\cite{SM}.}
    \label{fig:expt_setup}
\end{figure*}

The plot highlights the performance threshold a QM must meet to provide a genuine advantage. For instance, when the optimal repeater configuration yields elementary links of 200\,km, then storage times, $T_S$, on the order of 1\,ms are required to bridge classical-communication delays. 
This in turn, requires the QM platform to feature a storage efficiency of at least $\eta_o \geq 65\%$, a coherence time of $T_2 \geq 3$\,ms, and support a multimode capacity of $M \geq 60$ modes. More details of the model and assumptions are provided in the SM~\cite{SM}. Our QM platform addresses the simultaneous requirements of multimode capacity, high efficiency, and millisecond-scale storage times, and provides a clear path toward achieving the performance needed for practical repeater operation.

\textit{Experiment—}As a memory platform, we utilize a 5\,ppm \textsuperscript{167}Er\textsuperscript{3+}:Y\textsubscript{2}SiO\textsubscript{5} crystal due to its long coherence properties~\cite{ErYSO_MagneticField_bottger-2009, ErYSO_B7T_Cohrenece_Rancic2018}.
Specifically, we focus on crystallographic site 1 (1536.4\,nm) with an optical depth of $\approx2$ and $T_2 = 806.1 \pm 38.3$\,\textmu s (see SM \cite{SM}). 
The CPPE protocol uses two adiabatically chirped pulses (CPs) to suppress the primary echo and enable on-demand storage and retrieval, with the chirp bandwidth optimized to cover the input spectrum while minimizing spontaneous-emission noise. 
The experimental setup and pulse sequence for CPPE-based storage are shown in Figs.~\ref{fig:expt_setup}(a) and \ref{fig:expt_setup}(b), respectively.
Further details of the setup and protocol are described in the End Matter.

The weak coherent probe pulses are prepared with a Lorentzian shape with a full width at half-maximum (FWHM) of 0.75\,\textmu s.
We demonstrate on-demand storage of weak coherent probe pulses for various storage times ranging from $300$\,\textmu s to 1\,ms, as shown in Fig.~\ref{fig:expt_setup}(c).
The estimated mean photon number of the input probe pulse $n_\mathrm{in}$ is approximately 720 before the crystal with $\eta_T=20$\% transmission to the output fiber off atomic resonance. As depicted in Fig.~\ref{fig:expt_setup}(d), the storage efficiency $\eta_\mathrm{M}$ exhibits a characteristic exponential decay with increasing storage time, yielding efficiencies of 10.36\% and 1.42\% for storage times of 300\,\textmu s and 1\,ms, respectively. 
The retrieved echo into the output fiber, thus, contains an estimated mean photon number of $n_\mathrm{out} = n_\mathrm{in}\,\eta_{\text{T}}\,\eta_{\text{M}} = 14$ and $n_\mathrm{out}=2$ at 300\,\textmu s and 1 ms storage times, respectively. Moreover, a noise level of approximately five photons per storage mode is observed in the present system, which constitutes the main limitation in reducing the mean photon number of the input probe pulses. This noise primarily arises from spontaneous emission of residual ions remaining in the excited state due to the finite efficiency of the CPs as $\pi$ pulses. Additionally, the retrieved echo exhibits a broadened Gaussian profile caused by laser wavelength instability, which can be mitigated by stabilizing the laser through cavity locking.
More details on characterization of the noise and pulse broadening are discussed in SM~\cite{SM}.

\begin{figure}
    \centering
    \includegraphics[width=\columnwidth]{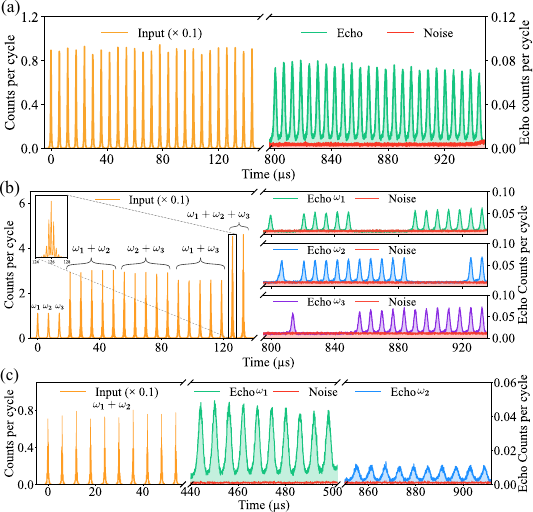}
    \caption{Demonstration of spectral-temporal multimode capacity.
    (a) A total of 25 temporal modes, with an average input mean photon number of $2489$, are stored for 800\,\textmu s and subsequently retrieved with an average efficiency of 2.67\% and SNR of 9.18. 
    (b) Spectral-temporal multimode capacity with selective recall. The left panel shows 20 temporal modes prepared using the three spectral modes ($\omega_1$, $\omega_2$, and $\omega_3$). For each spectral mode, 13 pulses are prepared, enabling all possible superpositions of the spectral modes within the 20 temporal modes. The inset shows the interference pattern obtained by superposition of the three spectral modes. All prepared modes are stored simultaneously for 800\,\textmu s, while each spectral mode is selectively recalled as shown in the right panels. The storage efficiency of the three spectral modes is 1.20\%, 1.64\%, and 1.66\%, respectively. 
    (c) Sequential recall of two spectral modes. The left panel shows a 10 weak coherent pulse input sequence prepared with two spectral modes ($\omega_1$ and $ \omega_2$). After simultaneous storage, they are recalled independently in the same cycle where $\omega_1$ is retrieved at 450\,\textmu s and $\omega_2$ is retrieved at 850\,\textmu s yielding efficiencies of 9.33\,\% and 2.08\,\%, respectively.}
    \label{fig:multimode}
\end{figure}

\textit{Multimode capacity—}The practicality of the memory can be significantly improved by storage of modes in different degrees of freedom, as discussed previously. 
In this section, we showcase the multimode capacity of temporal as well as spectral degrees of freedom simultaneously. 
First, we demonstrate storage of a train of 25 temporal modes for a duration of 800\,\textmu s as shown in Fig.~\ref{fig:multimode}(a). 
Each mode has a FWHM of 0.75\,\textmu s, and adjacent modes are separated by 7\,\textmu s from each other.
The maximum number of temporal modes, with a defined memory bandwidth, is mainly constrained by the duration $\tau_2$ as all the stored modes need to be recalled before the echo of the CP happening after $\tau_2$ from the second CP (see Fig. S5).

Next, we demonstrate storage and recall of three spectral modes, each stored in distinct spectral regions of the inhomogeneous broadening and serving as independent memory cells.
We first characterize the system and measure negligible crosstalk between adjacent cells separated by 4\,MHz when storing a 0.75\,\textmu s wide Lorentzian input pulse~\cite{SM}. 
To demonstrate spectrotemporal multimode operation, we prepare a train of $20$ temporal modes with various combinations of three spectral modes at frequencies $\omega_1$, $\omega_2$, and $\omega_3$, as shown in the left panel of Fig.~\ref{fig:multimode}(b). 
The detuning between each spectral mode is set to be 4\,MHz for negligible crosstalk.
This scenario is analogous to that of probabilistic entangled photon-pair sources, where photon pairs are simultaneously generated over a broad bandwidth \cite{EPS_wang-2021}.
Three consecutive CPs for each spectral mode are applied to enable storage of all the modes. 
Finally, the selective recall of only the desired spectral mode is performed by applying the corresponding second CP. The recalled output along with the noise for the three spectral modes is shown in the right panels of Fig.~\ref{fig:multimode}(b). This approach enables simultaneous storage of up to $20\times3 = 60$ modes. 

Finally, we demonstrate storage of two spectral modes and their sequential recall at different storage times, as shown in Fig.~\ref{fig:multimode}(c). 
A multitone train of ten input pulses is prepared with frequencies $\omega_1$ and $\omega_2$. 
Both the spectral modes are stored by applying two simultaneous CPs followed by recall of $\omega_1$ mode at 450\,\textmu s and recall of $\omega_2$ mode at 850 \,\textmu s.
This capability enables the utilization of all the spectral-temporal modes for establishing entanglement across neighboring links in a QR.
The spectral multiplexing comes at the cost of the cumulative noise from the spontaneous emission of all active memory cells, as shown by the red noise curve in Fig.~\ref{fig:multimode}. Some strategies to mitigate the noise are discussed in the following section. 

\begin{figure}
    \centering
    \includegraphics[width=\linewidth]{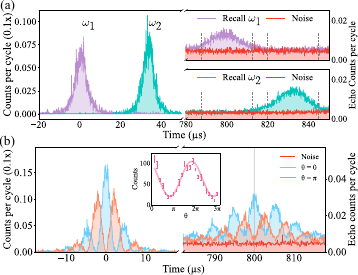}
    \caption{State characterization of stored spectral modes. (a) Two spectral modes with temporal width 10\,\textmu s, mean photon number $\approx 850$ and frequency detuning $\delta = \omega_1 - \omega_2 = 0.2$ MHz are prepared with a relative temporal offset (left). Each mode is individually recalled after a storage time of 800\,\textmu s and state overlap in $Z$ basis is estimated using the indicated temporal bins (right). 
    (b) The two modes are prepared in a superposition state with a relative phase $\theta$, producing an interference pattern (left). Both modes are simultaneously recalled after 800\,\textmu s (right). Echo counts within the shaded gray region are recorded as a function of $\theta$ (inset), yielding an interference fringe with a visibility of $75.5 \pm 0.6\%$.}
    \label{fig:fidelity}
\end{figure}

Another key performance metric of the memory is the overlap of the stored and retrieved states. To evaluate this, we store two spectral modes representing a frequency-bin qubit. 
The overlap in the rectilinear ($Z$) basis is determined by recalling each spectral mode individually after a storage time of 800\,\textmu s, as shown in Fig.~\ref{fig:fidelity}(a), yielding an average overlap of $\mathcal{F}_Z = 70.2\%$. 
To probe coherence, the two spectral modes are prepared in a superposition state with a relative phase~$\theta$ applied via the AOM, producing the interference pattern shown in Fig.~\ref{fig:fidelity}(b). 
The retrieved signal exhibits a similar interference pattern, confirming preservation of phase coherence during storage. 
By sweeping $\theta$, we obtain the interference visibility, corresponding to an $X$-basis overlap of $\mathcal{F}_X = 87.7\%$. 
The overall overlap of the retrieved state, averaged over mutually unbiased bases, is therefore estimated as $\mathcal{F} = 81.8\%$. 
Additional details of the measurement are provided in the SM~\cite{SM}.


\textit{Conclusion—}We demonstrate an Er\textsuperscript{3+} based telecom memory that integrates the key functionalities required for a practical quantum repeater. 
The memory supports storage times up to $1$\,ms with on-demand retrieval, alongside the capability of selective retrieval of multiple spectral modes. 
In more quantifiable terms, this corresponds to a memory coherence time $T_2 = 858.4$\,\textmu s and initial efficiency $\eta_o = 23.05 \%$. At the same time, we demonstrated spectral-temporal multimode capacity with $3\times20$ modes. Although these parameters do not allow a multiplexed repeater to surpass direct transmission in terms of the probability of entanglement distribution, realistic improvements to $T_2$ and $\eta_o$ for the demonstrated degree of multiplexing, $M=60$, would suffice (see Fig.~S2(a) in \cite{SM}). Specifically, the $\eta_o=65\%$ and $T_2=3$\,ms memory parameters generating the curves in Fig.~\ref{fig:Qrepeater}(b), have been achieved experimentally with rare-earth-doped Y\textsubscript{2}SiO\textsubscript{5}.

For this platform to become suitable for quantum repeater applications, noise remains the primary limitation, particularly for the storage of low mean photon number states. 
A straightforward strategy for mitigating this noise is to reduce the beam area through the crystal, thereby interacting with a smaller number of ions. 
This reduces the noise contribution while preserving the same optical depth and, consequently, the storage efficiency. 
A smaller collection mode can also suppress noise arising from spontaneous emission, either by reducing the diameter of the collection optics placed after the crystal or by implementing laser-written waveguides~\cite{ErYSO_AFC_Liu2022}. 
We estimate that this approach can reduce the noise by at least a factor of 50~\cite{SM}, making the storage of nonclassical light feasible.
The improved optical confinement would also enhance the Rabi frequency of the CPs~\cite{ErTFLN_Spectroscopy_wang-2022}, allowing for increased operational bandwidth of the memory. 
Recent studies in Er\textsuperscript{3+}:Y\textsubscript{2}SiO\textsubscript{5} have demonstrated both the identification and selective addressing of spin levels~\cite{ErYSO_Spectro_Stuart2021, ErYSO_Spectro_Kukharchyk2021, ErYSO_Zefoz_matsuura-2026}, fulfilling key prerequisites for the implementation of the noiseless photon echo (NLPE) protocol~\cite{QMem_NLPE_ma-2021}. 
Consequently, NLPE represents a realistic approach in this platform, providing a pathway to suppress spontaneous emission noise while enabling extended storage times through a four-level scheme.
Furthermore, a cavity under critical coupling can enhance efficiency toward unity~\cite{ErYSO_NanoBeam_craiciu-2019, REI_Cavity_meng-2026}, while a stronger magnetic field extends the coherence time to 4~ms \cite{ErYSO_MagneticField_bottger-2009}, enabling QRs that outperform direct transmission~\cite{SM}. In conclusion, the proposed improvements position the Er\textsuperscript{3+}:Y\textsubscript{2}SiO\textsubscript{5} platform as a strong candidate for building the hardware required for scalable, long-range quantum networks.

\textit{Data availability}—The data and scripts that support the findings of this article are publicly available~\cite{Dataset}.

\textit{Acknowledgments}—A.S. would like to thank Alberta Innovates for funding through the Graduate Student Scholarship. This work was supported by the Government of Alberta Major Innovation Fund Project on Quantum Technologies, the National Research Council of Canada through the Small Teams Initiative QPIC, the Department of National Defense through the Innovation for Defence Excellence and Security (IDEaS) program, the Canada Foundation for Innovation John R. Evans Leaders Fund (CFI-JELF), the Natural Sciences and the Engineering Research Council of Canada (NSERC) through the Discovery Grant program. 

\textit{Author Contribution}—A.S. implemented the experiment and analyzed the data with the help from N.G.K. A.S. and N.G.K. drafted the manuscript. D.O. supervised the project. All authors discussed the experimental procedures and results and edited the manuscript.

\section{End Matter}
 
The crystal is housed in a pumped He-3 cryostat, maintained at a temperature of approximately 500\,mK. 
The light is coupled into the crystal using single-mode fibers with pigtailed ferrules and gradient index lenses at both ends of the crystal. 
The components are selfaligned using a V-groove assembly, resulting in a total optical transmission of approximately $\eta_\mathrm{T} =20\%$. 
This compact and simple setup eliminates the need for costly nanopositioners or complex free-space setups, thus, simplifying the implementation for large-scale quantum networks. 
A vector magnet applies a static magnetic field of 3\,T oriented at an angle of $135^\circ$ relative to the $D_1$ axis along the $D_1$-$D_2$ plane of the crystal.

In our experimental setup, as depicted in Fig.~\ref{fig:expt_setup}(a), an external cavity diode laser and two AOMs are used to set the laser frequency to the peak absorption of the site~1 transition at 1536.056\,nm under the applied magnetic field.
The AOMs generate both the weak coherent input probe pulses and the strong chirped pulses needed for the CPPE protocol. 
After transmission from the crystal, two additional AOMs suppress all light except the input and echo signals, preventing saturation of the SNSPD by the chirped pulses. 
The AOMs also serve as frequency shifters to match the frequency of retrieved spectral modes in the described QR scheme in Fig.~\ref{fig:Qrepeater}(a).

The CPPE memory protocol sequence is depicted in Fig.~\ref{fig:expt_setup}(b). To ensure consistent performance, each storage cycle begins with a 50\,ms optical pumping step. This process resets the atomic population from the previous cycle, thereby avoiding a decrease in storage efficiency. 
The pumping transfers population from the surrounding region into the memory region, creating spectral pits approximately 5\,MHz wide on either side of the memory window. 
Following this, we include a 100\,ms wait time to allow for the spontaneous decay of ions back to the ground state manifold, as the optical lifetime is measured to be about 10\,ms~\cite{SM}.
Next, the weak input probe pulse is prepared at $t = 0$. 
Following a delay of $\tau_1$, the first chirped pulse with a duration of $\tau_{CP}$ is applied to adiabatically flip the population of the atomic ensemble. The CP pulse, $P(t) = A(t)~\sin[\phi(t)]$, has a sech hyperbolic envelope combined with a linear frequency sweep: 

\begin{equation}
    A(t) = A_0~\mathrm{sech} \left(\frac{10 t}{\tau_{CP}} \right),~~~
    \frac{d\phi}{dt} = \omega(t) = \omega_0 + \frac{2t\Delta}{\tau_{CP}}.
\end{equation}

The adiabatic chirp of this pulse imprints a detuning-dependent phase onto the atoms, which suppresses the primary echo~\cite{ChirpPulse_Adia_doll-2013, QMem_CPPI_morton-2022, QMem_CPPI_kamel-2025}.
Moreover, the adiabatic nature of the population transfer makes CPPE robust to power and frequency fluctuations~\cite{ChirpPulse_Adia_doll-2013, QMem_CPPI_kamel-2025}.
The span of the frequency chirp ($\Delta$) is optimized to cover the entire spectrum of the input pulse while avoiding the excitation of any additional ions, thereby reducing noise from spontaneous emission~\cite{ChirpPulse_ledingham-2010, ChirpPulse_duda-2023}. 
To retrieve the stored input modes, an identical second CP is applied after a delay of $\tau_2$. 
The atomic ensemble subsequently reemits the echo of the input mode after a duration of $\tau_2 - \tau_1$, resulting in a total storage time of $T_S = 2(\tau_2 + \tau_{CP})$. 

Although the protocol is designed to suppress the primary probe echo, imperfections in the CP can result in an ``unsilenced" echo being present. 
To prevent this unwanted echo from overlapping with the desired output, we set the timing of CP such that $2\tau_1 < \tau_2$. 
Additionally, the protocol results in an echo from the CPs themselves occurring after the recall of input modes, which is suppressed by a second pair of AOMs. More details on protocol and optimization of CPs are described in SM~\cite{SM}.


\bibliographystyle{apsrev4-2}
\bibliography{references}

\end{document}


\title{Supplemental Material: On-Demand Millisecond Storage of Spectrotemporal Modes at Telecom Wavelength}

\author{Anuj Sethia\,\orcidlink{0000-0002-7140-3662
}}
\author{Nasser Gohari Kamel\,\orcidlink{0009-0003-7869-2561}}
\author{Daniel Oblak\,\orcidlink{0000-0002-0277-3360}}

\email[Corresponding email: ]{doblak@ucalgary.ca}
\affiliation{Department of Physics and Astronomy, University of Calgary, Calgary, Alberta T2N 1N4, Canada.}
\affiliation{Institute of Quantum Science and Technology, University of Calgary, Calgary, Alberta T2N 1N4, Canada.}

\date{\today}

\maketitle

\section{Spectral properties} \label{sec:spectral_properties}

\begin{figure}[ht!]
    \centering
    \includegraphics[width=0.58\linewidth]{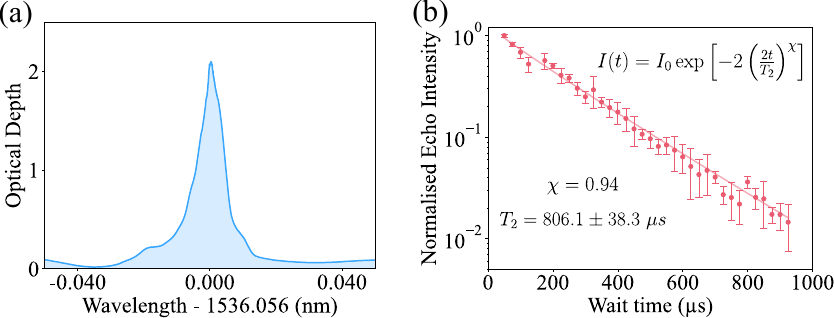}
    \caption{(a) Absorption spectrum of site 1 at 500\,mK with $B=3$\,T along $135^\circ$ from the $D_1$ axis in the $D_1-D_2$ plane. 
    (b) Optical coherence time ($T_2$) of site 1 ions using the two-pulse photon echo technique. The decay of the echo intensity is fitted with Mims expression \cite{ErYSO_Decoherence_Bottger2006}, resulting in $T_2 = 806.1 \pm 38.3$\,\textmu s and minimal spectral diffusion as $\chi = 0.94$.}
    \label{fig:spectrum}
\end{figure}

We utilize isotopically purified \textsuperscript{167}Er\textsuperscript{3+} ions doped in a Y\textsubscript{2}SiO\textsubscript{5} crystal of dimension $2\times 2\times 10$\,mm along the $D_1$, $D_2$, and $b$ axes, respectively.
The isotopic purification reduces the complexity of hyperfine levels in the absorption profile, while the low doping concentration suppresses ion-ion interactions, thereby enhancing the optical coherence time. 
The Er\textsuperscript{3+} ions occupy two crystallographic locations, replacing Y\textsuperscript{3+} ions in the Y\textsubscript{2}SiO\textsubscript{5} crystal. 
The two locations are spectrally distinguishable and have traditionally been labeled site~1 ($1536.4\,nm$) and site~2 ($1538.9\,nm$)~\cite{ErYSO_Hyperfine_Bottger2006}.
The inhomogeneously broadened absorption profile for site~1 is shown in Fig.~\ref{fig:spectrum}(a). 
The narrow inhomogeneous broadening is attributed to the low-strain environment experienced by Er\textsuperscript{3+} ions due to their ionic radius being similar to that of Y\textsuperscript{3+}. 
The absorption profiles also consist of two magnetically inequivalent subclasses of sites for each crystallographic site, which will split if a magnetic field is applied in an arbitrary direction. 
However, in cases of a magnetic field along the $b$ axis or in the $D_1-D_2$ plane, all of the subclasses of a given site become magnetically equivalent~\cite{ErYSO_MagneticField_bottger-2009}. 
Additionally, maximum optical coherence time is observed when the magnetic field is applied at an angle of $135^\circ$ from the $D_1$ axis along the $D_1-D_2$ plane of the crystal~\cite{ErYSO_MagneticField_bottger-2009}. 
We measure the optical coherence time in this configuration using the two-pulse photon echo decay technique as shown in Fig.~\ref{fig:spectrum}(b). 

\section{Entanglement Distribution Rates} \label{sec:repeater}
Entanglement distribution in a quantum repeater (QR) relies on performing entanglement swapping between entangled photon pairs generated at different nodes \cite{QRepeater_simon-2007, QRepeater_sangouard-2011}.
The process involves two stages. 
In the first stage, entanglement swapping is carried out between the spectrotemporally multiplexed idler photons from adjacent nodes that propagate through the elementary links and interfere on a beam splitter as shown in Fig.~1(a). They are further spectrally demultiplexed, e.g., using a virtually imaged phased array (VIPA), as recently demonstrated by Chakraborty \textit{et al.} \cite{QRepeater_Multimode_chakraborty-2025}.
Once all elementary links have successfully completed this initial swapping, a second swapping operation is performed between the signal photons that had been stored and are now simultaneously retrieved from the quantum memories (QMs).
Therefore, the overall entanglement distribution success probability, $P_{s}$, for a total link length of $L$ divided into $n_l$ elementary links is determined by the combined success probabilities of these two swapping operations, expressed as \cite{REI_Multimode_sinclair-2014, REI_Multimode_lago-rivera-2021}:

\begin{equation}
   P_{s} = \left[1-\left(1-\frac{1}{2}\left(\rho ~ 10^{-\alpha  L /2n_l} ~ \eta_{D}^{i}\right)^\beta \right)^{M_s M_t}\right]^{n_l}  \frac{ (\eta_{D}^{s} ~\eta_{M}^{s})^{2n_l}}{2^{n_l-1}}, \quad \eta_{M}^{\textit{s}} = \eta_o \exp \left[-\frac{4L}{n_l v T_2}\right].  \label{equ:rate}
\end{equation}

\begin{figure*}
    \centering
    \includegraphics[width=\linewidth]{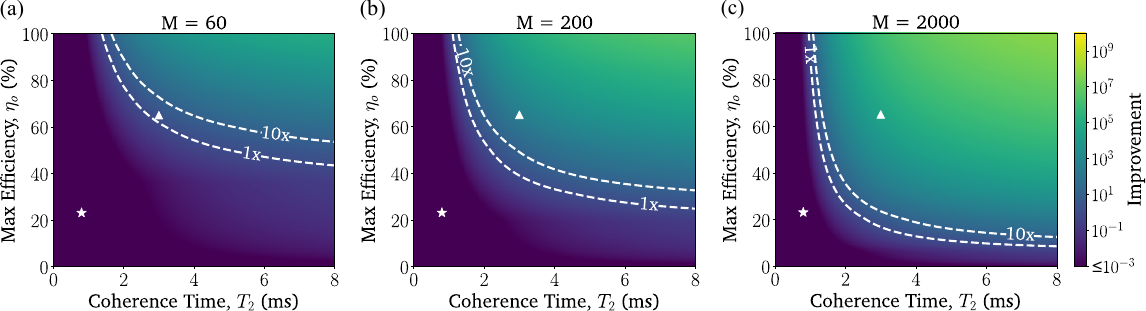}
    \caption{The ratio of success probabilities between a QR and direct fiber transmission is shown for a total link length of L = 500 km, plotted as a function of coherence time ($T_2$) and maximum efficiency ($\eta_o$). Each subplot corresponds to a different multimode capacity: (a) M = 60, (b) M = 200, (c) M = 2000. The star indicates the performance of the QM demonstrated in this work ($T_2 = 806.1$\,\textmu s and $\eta_o = 23.05\%$), while the triangle marks the achievable quantum memory performance ($T_2 = 3$ ms \cite{ErYSO_MagneticField_bottger-2009} and $\eta_o = 65\%$~\cite{REI_Cavity_duranti-2024}) to surpass the direct transmission benchmark—achievable through cavity enhancement and stronger magnetic fields.}
    \label{fig:RepeaterPerf}
\end{figure*}

\begin{table}
    \centering
    \begin{tabular}{ccl}
        \textbf{Parameter} & \textbf{Value} & \textbf{Description} \\ \hline \hline
        $\rho$ & 0.9 & Entangled pair generation probability \\
        $\alpha$ & 0.21 & SMF-28 loss coefficient (dB/km) \\
        $\beta$ & 2 & Exponent for detection scheme \\
        $\eta^i_D$ & 0.9 & Idler detection efficiency \\
        $\eta^s_D$ & 0.9 & Signal detection efficiency \\
        $M_t$ & 20 & Temporal multimode capacity \\
        $M_s$ & 3, 10, 100 & Spectral multimode capacity \\
        $\eta^s_M$ & - & QM storage efficiency \\
        $\eta_o$ & 65\% & QM's maximum efficiency \\
        $T_2$ & 3 ms & QM's coherence time \\
        $\nu$ & $c/1.468$ & Speed of light in SMF-28 fiber \\ \hline
    \end{tabular}
    \caption{Parameters and values used for estimation of $P_{s}$.}
    \label{tab:QRepeater_params}
\end{table}

The first term in Eq.~(\ref{equ:rate}) represents the probability that at least one mode, among the total of $M_s \times M_t$ available modes, successfully undergoes entanglement swapping across each of the $n_l$ elementary links.
Unlike in Refs.~\cite{REI_Multimode_sinclair-2014, REI_Multimode_lago-rivera-2021}, the temporal multimode capacity appears in the exponent due to the on-demand recall capability of our QM, which enables interference between arbitrary temporal modes across adjacent links.
The second term accounts for the probability of a bell-state measurement (BSM) between the signal photons after they are retrieved from the QMs.
The parameters used for estimating $P_{s}$ in Fig.~1(b) in the main text are summarized in Table~\ref{tab:QRepeater_params}. 
The entanglement distribution rate, $R$, is given by $R = \nu P_{s}$ where $\nu$ is the repetition rate of the memory storage cycle, accounting for memory preparation steps.
Fig.~\ref{fig:RepeaterPerf} illustrates the entanglement distribution improvement using the multiplexed QR, specifically the ratio of success probabilities over direct fiber transmission. 
This improvement is analyzed based on three key QM metrics: coherence time ($T_2$), maximum efficiency ($\eta_o$), and multimode capacity ($M$).
The plots also pinpoint the performance requirement for the QM necessary to surpass the success probability benchmark set by direct fiber transmission, demonstrating where the QR offers a tangible advantage.

\section{Chirp Pulses} \label{sec:ChirpPulse}
The chirped pulses (CPs) play a crucial role in determining the performance of the storage protocol, as their primary function is to adiabatically and completely transfer population between the ground and excited states and, for the CPPE protocol, imprint a spectrally dependent phase to inhibit/silence a primary echo~\cite{ChirpPulse_Adia_Hanamura_2026}. The effectiveness of the population transfer, in turn, affects both the storage efficiency and the noise arising from residual population in the excited state at the time of the secondary echo.

To optimize the storage performance, we systematically varied the CP parameters, including the power, duration, and chirp range, across several candidate pulse configurations. The optimization aimed to maximize the storage efficiency while minimizing the spontaneous emission noise, which serve as indicators of the adiabaticity and completeness of the population transfer, respectively. Among the pulse shapes investigated, promising performance was obtained using a hyperbolic secant (sech) envelope combined with either a hyperbolic frequency sweep (ROSE) or a linear frequency sweep (CPPE). We find that the CPPE configuration provides the best overall balance between the two metrics.

\begin{figure}
    \centering
    \includegraphics[width=\linewidth]{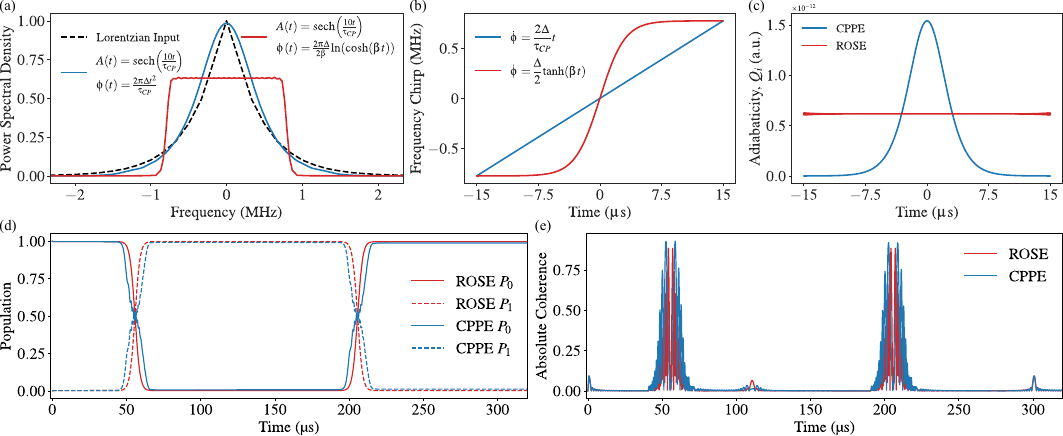}
    \caption{Comparison of the ROSE and CPPE CPs and their simulated dynamics. (a) Power spectral density of CPs compared with the Lorentzian input pulse (FWHM = 750~ns). The CPs are assumed to have identical amplitude, duration of 30\,\textmu s, and chirp bandwidth $\Delta = 1.5$~MHz. 
    (b) Frequency chirp profiles of the different pulse configurations. 
    (c)  Instantaneous adiabaticity parameter $Q_i$ of the CPs over their respective pulse durations.
    (d) Simulated population dynamics for the ROSE and CPPE protocols, showing the evolution of the ground ($P_0$) and excited ($P_1$) state populations over time. 
    (e) Corresponding absolute coherence as a function of time for both protocols.}
    \label{fig:CP_fft}
\end{figure}

The superior performance of CPPE can be understood from the spectral and adiabatic properties of the CPs. Fig.~\ref{fig:CP_fft}(a) shows the power spectral density of the two optimized pulse configurations. Compared with the CPPE pulse (blue curve), the ROSE pulse (red curve) exhibits a lower peak spectral density within the chirped bandwidth because its power is distributed more uniformly over the frequency sweep. Fig.~\ref{fig:CP_fft}(b) compares the corresponding frequency chirp profiles, highlighting the significantly higher instantaneous chirp rate of the ROSE pulse at the center of its range, which reduces the phase imprint and the population transfer in the range overlapping most with the input probe pulse. 

Conventionally, the adiabaticity of CPs is characterized by $Q \approx \Omega_R^2/k$~\cite{ChirpPulse_Adia_doll-2013,QMem_CPPI_morton-2022}, where $\Omega_R$ is the Rabi frequency and $k=\Delta/\tau_{\mathrm{CP}}$ is the chirp rate. However, for non-uniform frequency sweeps, a more appropriate metric is the instantaneous adiabaticity,
\begin{equation}
    Q_i(t)=\frac{\Omega_R^2(t)}{\left|\frac{d\Delta(t)}{dt}\right|}.
\end{equation}
As shown in Fig.~\ref{fig:CP_fft}(c), the CPPE pulses exhibit higher instantaneous adiabaticity in the central spectral region overlapping most with the probe pulse, whereas the hyperbolic chirp profile used in ROSE maintains a relatively constant adiabaticity throughout the pulse duration. This enhanced adiabaticity, resulting from the slower chirp rate of the CPPE pulse within the relevant spectral region, improves the phase imprint, leading to more effective suppression of the undesired echo and enhanced retrieval of the desired echo~\cite{ChirpPulse_Adia_Hanamura_2026}. The tradeoff is that the CPPE pulse less efficiently transfers population at the start and the end of the chirp range, which, though not overlapping much with the probe pulse spectrum, does have the potential to introduce spontaneous emission noise from atoms remaining in the excited state during the emission of the secondary echo.

To further understand the population dynamics, we numerically investigate the ensemble evolution by solving the Hamiltonian of the system~\cite{QMem_CPPI_kamel-2025} under the weak-probe approximation. Due to the computational complexity of the simulations, we restrict the model to an ensemble of 1500 ions and neglect homogeneous dephasing and spontaneous decay processes.

As shown in Fig.~\ref{fig:CP_fft}(d), the ROSE protocol achieves more complete population transfer over the whole spectral range at the end of the sequence, resulting in a lower residual excited-state population at the time of the secondary echo. However, this improved overall population transfer is accompanied by a higher undesired primary echo, as indicated by the residual coherence at $t\approx 110$\,\textmu s in Fig.~\ref{fig:CP_fft}(e). Despite these differences, both protocols produce comparable amplitudes for the desired echo in the idealized simulations. Experimentally, however, we observe a reduced echo amplitude for the ROSE protocol compared to CPPE. 

We conclude that CPPE pulses provide a better balance between adiabaticity and population transfer efficiency, resulting in superior performance within our optimization framework. More generally, alternative CPs may further improve this balance and provide additional opportunities for optimizing the storage process. Additionally, more comprehensive simulations of the population dynamics, including effects such as dephasing, inhomogeneous coupling~\cite{ChirpPulse_Adia_Hanamura_2026} and pulse propagation, will be valuable for gaining deeper insight into the performance of the protocols.

\section{Noise Analysis} \label{sec:Noise}

\begin{figure}[htb!]
    \centering
    \includegraphics[width=0.6\linewidth]{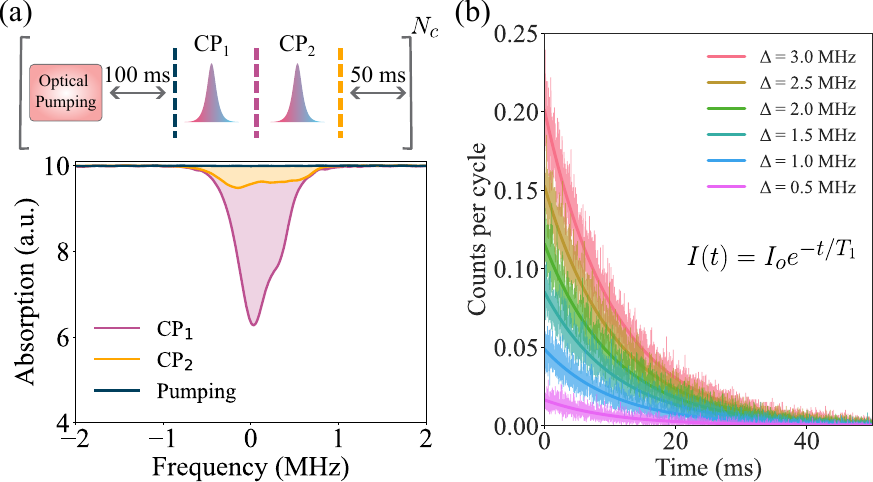}
    \caption{(a) Absorption spectra recorded at successive stages of the CPPE protocol cycle, showing the spectrum after optical pumping, after the first chirped pulse (CP\textsubscript{1}), and after the second chirped pulse (CP\textsubscript{2}). 
    (b) Spontaneous emission of ions following the application of a CP with different chirp ranges ($\Delta$), while keeping the ratio of chirp range to power constant. The decay exhibits the characteristic exponential behavior with a decay constant of $T_1 = 10.68 \pm 0.07$\,ms for all chirp ranges.}
    \label{fig:noise}
\end{figure}

The primary source of noise observed at the expected echo originates from the spontaneous emission of ions remaining in the excited state.
This decay process begins after the application of the first CP and continues even after the second CP, as the latter cannot fully transfer the excited population back to the ground state.
To evaluate the effectiveness of the CPs, we  measured absorption spectrum at various stages of the storage protocol as shown in Fig.~\ref{fig:noise}(a). 
We estimate the residual excited-state population after the second chirp pulse (yellow shaded region) to be approximately $18\%$ of the population initially excited by the first pulse. The decay of these ions contributes to a noise floor of about five photons during the expected echo, assuming a 6\,\textmu s retrieval window and a memory bandwidth of 1.5~MHz. Furthermore, we validate the scaling of the noise level by increasing the number of ions spectrally addressed by the CPs. 
As expected, the noise level scales linearly with the chirp range ($\Delta$) of the CP, as illustrated in Fig.~\ref{fig:noise}(b).

At the current noise level, operation in the single-photon regime (mean photon number $\approx 0.1$) is not yet feasible, and achieving this benchmark will require suppressing the noise by at least two orders of magnitude. Fortunately, there are two straightforward and complementary approaches to mitigate the noise.

The first is to increase the optical mode confinement within the crystal. The number of ions participating in the storage process scales with the interaction volume and therefore $N_{\mathrm{ions}} \propto D^2$, where $D$ is the optical mode diameter. Hence, reducing the mode size employs fewer ions in the storage and therefore reduces the spontaneous emission. Importantly, the storage efficiency is governed primarily by the optical depth rather than the total number of addressed ions and is therefore expected to remain largely unchanged. 

Furthermore, for a fixed optical power $P$, the Rabi frequency is inversely proportional to the optical mode diameter,

\begin{equation}
\Omega_R=\frac{\mu_d}{\hbar D}\sqrt{\frac{8P}{\pi c\epsilon_0}}
\propto D^{-1},
\end{equation}

where $\mu_d$ is the transition dipole moment. Consequently, tighter optical confinement increases the achievable Rabi frequency, enabling operation with a larger memory bandwidth and potentially improving the overall performance of the storage protocol. Such enhanced mode confinement can be achieved using a smaller gradient index lenses with mode diameters as small as $D=0.25$~mm, which are expected to reduce the noise by a factor of $\approx 50$. Even stronger confinement is possible using laser-written waveguides, which support low-loss optical modes with diameters of approximately 10\,\textmu m~\cite{ErYSO_AFC_Liu2022}. While these waveguides are expected to further suppress the noise, a quantitative estimate requires detailed simulations of the coupling of spontaneous emission into the guided optical mode.

A second approach, which is fully complementary to the first, is to reduce the fraction of spontaneous emission collected by the detection optics. Since spontaneous emission is emitted over a wide range of angles, reducing the collection aperture rejects this unwanted fluorescence while still efficiently collecting the retrieved memory mode. To evaluate this effect, we estimate the geometric collection efficiency for a 10~mm long crystal, assuming an average emitter distance of $z \approx 5$~mm from the collection lens. The external collection angle is given by $\theta_{\mathrm{ext}}=\tan^{-1}(r/z)$, where $r$ is the lens radius. The corresponding internal collection angle is determined by Snell's law, $n \sin \theta_{\mathrm{int}} = \sin \theta_{\mathrm{ext}}$ with n=1.8 for $\mathrm{Y_2SiO_5}$. The fraction of spontaneous emission collected is then

\begin{equation}
\rho=\frac{\Omega}{4\pi}=\frac{1}{2}\left(1-\cos\theta_{\mathrm{int}}\right).
\end{equation}

Comparing collection lenses with diameters of 1.8~mm and 0.25~mm yields

\begin{equation}
\frac{\rho(D=0.25~\mathrm{mm})}{\rho(D=1.8~\mathrm{mm})}\approx\frac{1}{50},
\end{equation}

indicating that only a small fraction of the spontaneous emission reaches the detector when a smaller collection lens is used.

In general, tighter optical confinement provides a clear pathway toward the noise suppression required for single-photon level operation. Beyond enhancing the signal-to-noise ratio for non-classical light storage, such implementations also offer an avenue toward device miniaturization. Such architectures could enable the integration of multiple memory elements on a single crystal, thereby substantially increasing the spatial and spectral multiplexing capacity of the memory.

\section{Temporal Jitter} \label{sec:jitter}

\begin{figure}[ht!]
    \centering
    \includegraphics[width=0.6\linewidth]{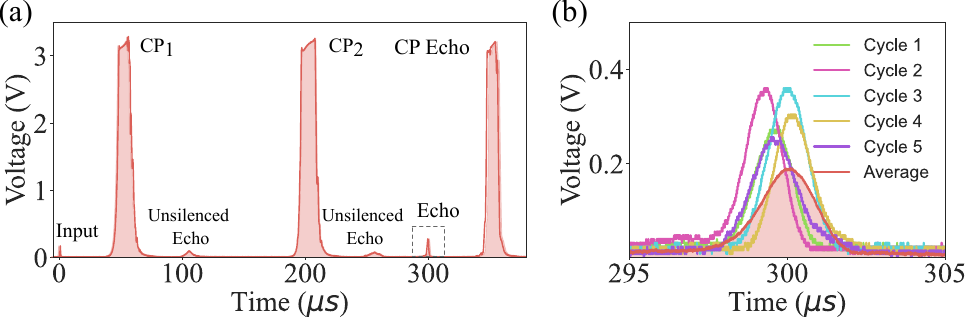}
    \caption{CPPE protocol implementation.
    (a) Strong input pulse (750\,ns) is stored for 300\,\textmu s, and the resulting trace is recorded using a photodiode. The CPs saturate the detector, with a small unsilenced echo visible immediately after the CPs. An additional CP-induced echo appears after the recalled echo and also saturates the detector.
    (b) The averaged echo over 50 traces (shown in red) is plotted along with individual traces. The observed variation in the echo recall time arises from laser frequency fluctuations. }
    \label{fig:jitter}
\end{figure}

To demonstrate that the broadening of the recalled echo is a direct consequence of the finite laser linewidth, a strong input signal is stored for 300\,\textmu s. 
The complete storage sequence is recorded using a photodetector, and the averaged trace is shown in Fig.~\ref{fig:jitter}(a). 
The trace depicts the two CPs saturating the photodetector and the CP-induced echo occurring at about 350\,\textmu s due to a two-pulse photon echo-like scheme. 
We also observe the ``unsilenced" echo after the CPs due to higher input intensity. 
Fig.~\ref{fig:jitter}(b) presents multiple individual recall events alongside the averaged signal.
The individual recalled pulses have an estimated FWHM of approximately  1.4\,\textmu s, while the averaged echo exhibits a FWHM of about 2\,\textmu s.
While each trace reproduces the input pulse, their precise retrieval times vary between storage cycles. 
This temporal jitter, originating from the laser's frequency instability, causes the averaged echo to appear significantly broadened, confirming that laser frequency uncertainty is the dominant mechanism responsible for this effect.

\section{Spectral crosstalk} \label{sec:cross_talk}

\begin{figure}[h!]
    \centering
    \includegraphics[width=\linewidth]{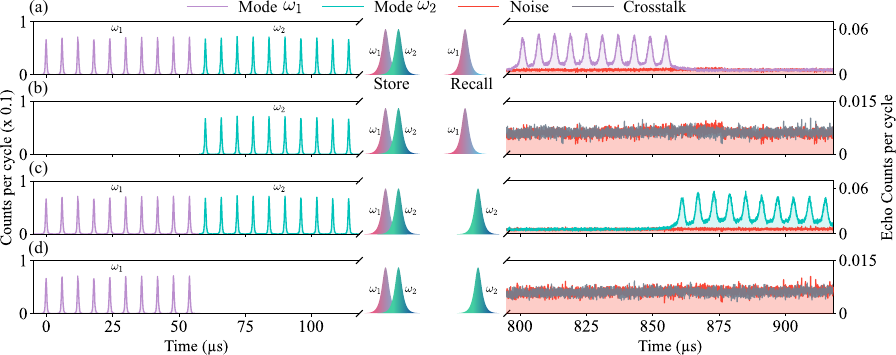}
    \caption{Measurement of spectral crosstalk between two adjacent memory cells in the spectrally multiplexed quantum memory. Subplots (a) and (c) show the storage and selective recall of frequencies $\omega_1$ and $\omega_2$. Subplots (b)
    and (d) present a comparison of the crosstalk signal with the background noise for each memory cell, respectively.}
    \label{fig:cross-talk}
\end{figure}

A key requirement for a spectrally multiplexed QM is that adjacent spectral channels operate independently, with negligible crosstalk between neighboring memory cells. To quantitatively assess the level of spectral crosstalk in our system, we prepare ten temporal modes each at two distinct frequencies, $\omega_1$ and $\omega_2$, separated by a detuning of 4 MHz, as illustrated in the left panels of Fig.~\ref{fig:cross-talk}. Both spectral modes are simultaneously stored by applying corresponding CPs precisely matched to each mode's center frequency, and then selectively retrieved one mode at a time (Fig.~\ref{fig:cross-talk}(a) and (c)).

To assess crosstalk from the $\omega_2$ cell into the memory cell $\omega_1$, we store only the mode $\omega_2$ and compare the retrieved noise to the background level obtained when no modes are stored (Fig.~\ref{fig:cross-talk}(b)). Conversely, crosstalk from $\omega_1$ into $\omega_2$ is evaluated by storing only the $\omega_1$ mode and performing the same comparison (Fig.~\ref{fig:cross-talk}(d)). In both cases, the measured difference in noise level is negligible, confirming that adjacent memory cells remain spectrally isolated and that the memory exhibits essentially no crosstalk between neighboring frequency channels.

\section{State Characterization} \label{sec:fidelity}
We characterize the memory performance by estimating the overlap between the stored and retrieved states, akin to the fidelity for qubit storage. The state is encoded in two spectral modes ($\omega_1$ and $\omega_2$) with detuning $\delta =$ 0.2~MHz representing frequency-bin qubit, defined as $\ket{0} \equiv \ket{\omega_1}$ and $\ket{1} \equiv \ket{\omega_2}$. For the rectilinear ($Z$) basis, the two spectral modes are prepared with a temporal separation of 35\,\textmu s (Fig.~4(a) for straightforward distinction without any spectral filtering. For the superposition ($X$) basis, both modes are prepared simultaneously with no temporal delay, forming the state

\begin{equation}
    \ket{\psi} = \frac{\ket{\omega_1} + e^{i\theta}\ket{\omega_2}}{\sqrt{2}}.
\end{equation}

The relative phase $\theta$ and frequency detuning $\delta = \omega_1 - \omega_2 = 0.2~\text{MHz}$ give rise to an interference pattern (Fig.~4(b). A broader temporal pulse width of 10\,\textmu s is chosen for this experiment to fully resolve the interference fringes. Following preparation, the spectral modes are stored in two independent memory cells using the corresponding CPs. For the $Z$ basis, selective recall is achieved by applying only the second CP associated with the desired mode, analogous to Fig.~\ref{fig:cross-talk}. However, for the $X$ basis, both modes are recalled simultaneously by applying CPs to both memory cells.

For the $Z$-basis measurement, the spectral modes are mapped onto distinct temporal modes, and the state overlap is determined from photon counts within 25\textmu s wide temporal bins, as indicated in right panel of Fig.~4(a). Let $N_1$ and $N_2$ denote the total counts in the first and second temporal bins, respectively, including noise. Thus, the overlap for recalling each state is given by

\begin{equation}
    \mathcal{F}_{1} = \frac{N_{1}}{N_1 + N_2} = 67.5\%; \quad \mathcal{F}_{2} = \frac{N_{2}}{N_1 + N_2} = 72.9\%,
\end{equation}

with the average $Z$-basis overlap as $\mathcal{F}_{Z} = (\mathcal{F}_{1} + \mathcal{F}_{2})/2 = 70.2\%$. 

For the $X$ basis, the state overlap is inferred from the visibility of the interference fringe obtained by sweeping the relative phase $\theta$. Photon counts are integrated within a narrow temporal window (gray region in Fig.~4(b)) to determine the visibility, $V = 75.5 \pm 0.6 \%$ via a sinusoidal fit. Thus, $X$-basis overlap is calculated as $\mathcal{F}_X = (1+V)/2 = 87.7\%$. This overlap is limited by both spontaneous emission noise and temporal jitter of the recalled echo (Sec.~\ref{sec:jitter}). Finally, the overall state overlap is estimated using

\begin{equation}
    \mathcal{F} = \frac{\mathcal{F}_{Z} + 2\mathcal{F}_X}{3} = 81.8\%, 
\end{equation}

which corresponds to the average overlap for a qubit encoded in mutually unbiased bases. Note that the state overlap in both bases is currently limited primarily by spontaneous emission noise. After subtracting the estimated noise contribution, we obtain overlaps of $\mathcal{F}_{Z} = 97.4\%$ and $\mathcal{F}_{X} = 91.1\%$. Implementing the noise-reduction strategies of Sec.~\ref{sec:Noise}, which are expected to suppress the noise by more than a factor of 50, would enable similarly high overlaps at low mean photon numbers.

\section{Memory Bandwidth} \label{sec:bandwidth}

\begin{table}[h!]
    \centering
    \begin{tabular}{cccccc}
         \textbf{Width (ns)} & \textbf{$\tau_{CP}$ (\textmu s)} & \textbf{$\Delta$ (MHz)} & \textbf{Efficiency (\%)} & \textbf{SNR} \\ \hline \hline
         750 & 30 & 1.5 & 8.49 &  5.52 \\
         500 & 40 & 2.2 & 6.01 & 4.11 \\
         250 & 60 & 4.5 & 1.62 & 1.42 \\ \hline
    \end{tabular}
    \caption{CP parameters with a peak power of 4 mW for different input pulse widths, along with the corresponding storage efficiency and SNR. The mean photon number of the input pulse is 513.}
    \label{tab:bandwidth}
\end{table}

An important performance metric for a QM is its storage bandwidth.
A larger bandwidth allows the storage of shorter input pulses, enabling more temporal modes and thereby facilitating efficient entanglement distribution.
In the CPPE protocol, the achievable bandwidth is primarily limited by the available optical power used to generate the chirped pulses required for broadband operation.
However, increasing the bandwidth also results in higher noise levels, since a larger number of ions are addressed by the chirped pulses.
Despite these trade-offs, we characterize the storage efficiency achievable for various input pulse widths at a fixed storage time of 500\,\textmu s.
The corresponding CP parameters, along with the measured QM efficiency and SNR, are reported in Table~\ref{tab:bandwidth}.

\section{Performance Comparison} \label{sec:performance}

\begin{figure}[h!]
    \centering
    \includegraphics[width=0.45\linewidth]{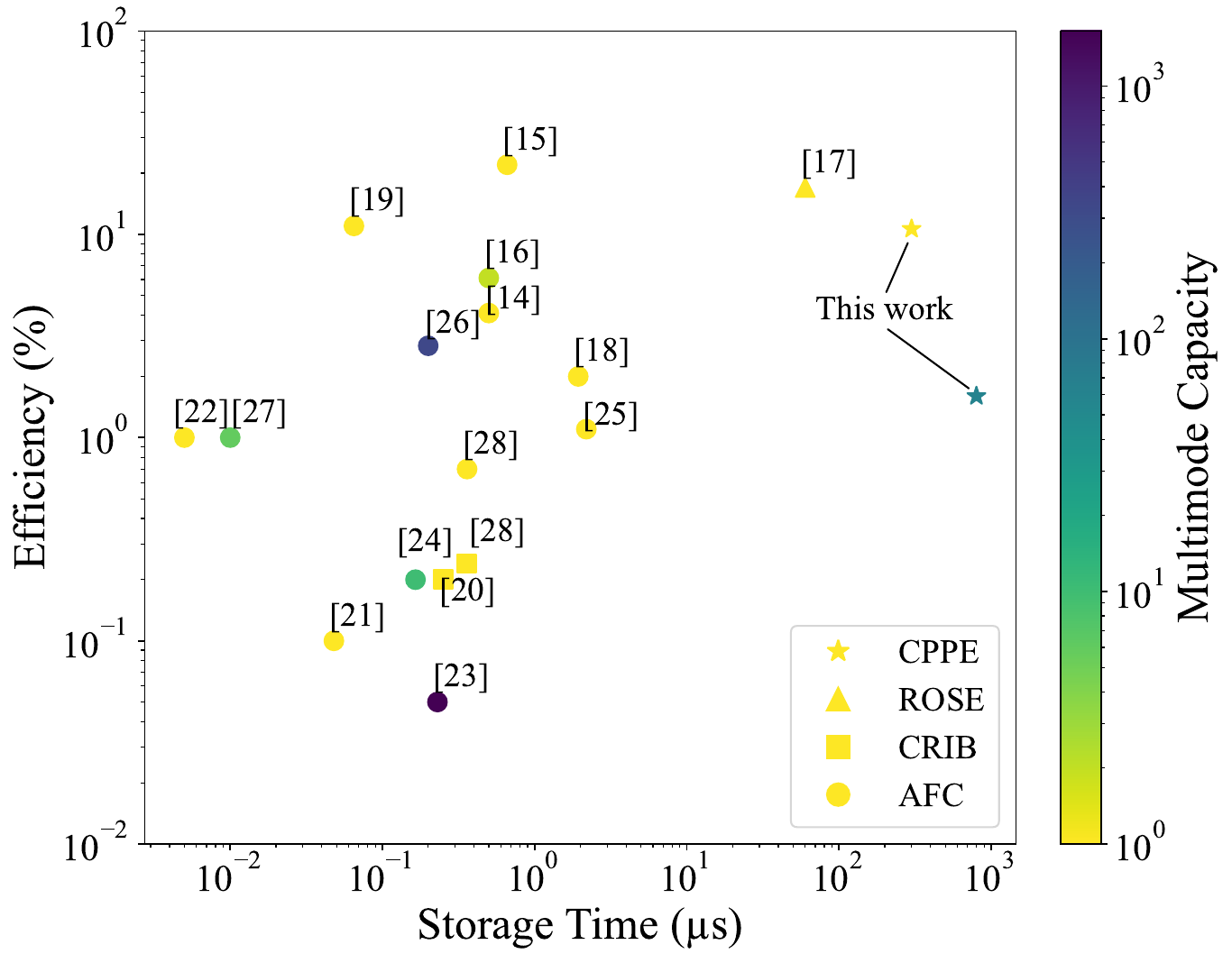}
    \caption{Performance comparison of erbium-based quantum memory demonstrations. Storage efficiency is plotted as a function of storage time for various implementations employing different protocols. The color scale indicates the reported multimode capacity for each demonstration. Data from Refs. \cite{ErYSO_AFC_Yasui2024,ErYSO_Spectro_Stuart2021,ErYSO_Multidimentional_Li2024,ErYSO_ROSE_minnegaliev-2023,ErYSO_EntStorage_jiang-2023,ErYSO_AFC_Liu2022,ErYSO_CRIB_Lauritzen2010,ErLN_AFC_Ent_askarani-2019,ErFiber_saglamyurek-2015,ErFiber_wei-2024,ErYSO_NanoBeam_craiciu-2019,ErYSO_Teleportation_an-2025,ErLN_Waveguide_zhang-2023,ErFiber_ENT_saglamyurek-2016,ErYSO_CRIB_AFC_lauritzen-2011}
    }
    \label{fig:performance}
\end{figure}

To benchmark the performance of our implementation, we compare the storage efficiency, storage time, and multimode capacity with erbium-based systems reported in the literature, as shown in Fig.~\ref{fig:performance}.

\bibliographystyle{apsrev4-2}
\bibliography{references}